\documentclass[a4paper]{jpconf}
\usepackage{amsmath}
\usepackage{amssymb}
\usepackage[square,sort&compress,numbers]{natbib}
\usepackage{upgreek}
\DeclareMathAlphabet{\eurm}{U}{eur}{m}{n}
\newcommand{\nasl}{{\rlap{\raise1pt\hbox{\,/}}\nabla}}
\newcommand{\dsl}{{\rlap{\raise1pt\hbox{/}}\mathrm{d}}}
\newcommand{\dLor}{\eurm{d}_{\scriptscriptstyle{\text{Lor}}}}
\newcommand{\gD}{\upgamma_{\!\scriptscriptstyle\upeta}}
\newcommand{\cev}[1]{\smash{\overset{\smash{{}_{\gets}}}{#1}}}
\newcommand{\lin}{{\scriptscriptstyle\bigstar}}
\newcommand{\idot}{{\scriptstyle\boldsymbol{\dot{}}}}
\newcommand{\txtm}[1]{\mbox{$\smash{#1}$}}
\begin{document}
\title{The distance formula in algebraic spacetime theories}

\author{D Canarutto and E Minguzzi}

\address{Dipartimento di Matematica e Informatica ``U. Dini'',\\ Universit\`a degli
Studi di Firenze,  Via S. Marta 3,  I-50139 Firenze, Italy}

\ead{daniel.canarutto@unifi.it, ettore.minguzzi@unifi.it}

\begin{abstract}
The Lorentzian distance formula, conjectured several years ago by Parfionov and Zapatrin, has been recently proved by the second author. In this work we focus on the derivation of an equivalent expression in terms of
the geometry of 2-spinors by using a partly original approach due to the first author. Our calculations clearly show the independence of the algebraic distance formula of the observer.
\end{abstract}

\section{Introduction}
The \emph{distance} between any two points of a Riemannian manifold $M$
can be expressed through \emph{Connes' formula}~\cite{connes95,parfionov00}: for any $x,y\in M$
\begin{equation}\tag{R1}\label{equation:Riemanniandistanceformula1}
\eurm{d}(x,y)=\sup_{\|\nabla\!f\|\leq1}\bigl\{f(y)-f(x)\bigr\}~,
\end{equation}
where the supremum is over the continuously differentiable 1-Lipschitz functions.
Moreover, as shown by Connes it admits the equivalent expression
\begin{equation}\tag{R2}\label{equation:Riemanniandistanceformula2}
\eurm{d}(x,y)=\sup_{f\in\mathcal{A}}\bigl\{|f(y)-f(x)|:\|[D,f]\|\leq1\bigr\}~,
\end{equation}
where the space \txtm{\mathcal{A}\equiv\mathrm{C}^1(M,\mathbb{R})}
is regarded as an algebra of operators (with pointwise multiplication)
and $D$ is the Dirac operator,
acting on sections of a spinor bundle
over $M$.
The latter expression is related to the so-called `algebraic' approach,
in which geometric information about an underlying manifold $M$
can be derived from a \emph{spectral triple}
\[ \bigl(\mathcal{A},\mathcal{H},D\bigr)~,\]
where $\mathcal{H}$ is a Hilbert space, $\mathcal{A}$ is an algebra of operators on it
and $D$ is a special operator.

Analogous results in Lorentzian geometry were established only recently. Let $(M,g)$ be a time-oriented Lorentzian manifold, namely a {\em spacetime}. Our choice for the Lorentzian metric signature is  $(+,-,-,-)$.
The Lorentzian distance
$\dLor(x,y)$
is defined on a spacetime  as
the supremum of the lengths of all piecewise $\mathrm{C}^1$,
future-oriented causal curves $\gamma$ joining $x$ and $y$
\[
\dLor(x,y)=\sup_{\gamma}\int_\gamma\sqrt{g(\dot \gamma,\dot \gamma)} \, d t.
\]
One sets \txtm{\dLor(x,y)=0} if $x$ and $y$ are not causally related \cite{beem96}.

The Lorentzian analog of~\eqref{equation:Riemanniandistanceformula1} reads
\begin{equation}\tag{L1}\label{equation:Lorentziandistanceformula1}
\dLor(x,y)=\inf_{f\in\mathcal{S}}\Bigl\{[f(x)-f(y)]^+\Bigr\}~,\qquad
[r]^+\equiv\max\{0,r\}\,,~r\in\mathbb{R}~,
\end{equation}
where $\mathcal{S}$ is the space of all \emph{steep} temporal functions,
i.e.\  $\mathrm{C}^1$ functions \txtm{M\to\mathbb{R}} such that $\mathrm{d}\!f$  fulfills
$\txtm{\langle\mathrm{d}\!f,{\scriptstyle Y}\rangle\geq\|{\scriptstyle Y}\|}\equiv \sqrt{g(\scriptstyle Y,\scriptstyle Y)}$
for every timelike future-oriented vector field ${\scriptstyle Y}$.

This formula was conjectured
by Parfionov\,\&\,Zapatrin~\cite{parfionov00},
used by Franco\,\&\,Eckstein~\cite{franco13} as an hypothesis
for their spectral triple formulation (see below and also Franco~\cite{franco18}),
and eventually proved by the second author~\cite{minguzzi17,minguzzi17d}
under the hypothesis of global hyperbolicity (and also by weaker assumptions). Although there were previous works devoted to the proof of \eqref{equation:Lorentziandistanceformula1}, they  failed to show that the steep functions in the formula could be taken to be defined all over of $M$, cf.\ \cite{moretti03} and $\mathrm{C}^1$, cf.\ \cite{franco10}.

The spectral triple reformulation of this formula, due to Franco and Eckstein~\cite{franco13}
 applies to spacetimes of arbitrary dimension
but is subjected to the choice of an observer.
In this work our aim is to obtain a spectral triple formulation in the 4-dimensional case, by using the formalism of 2-spinors. Our approach has the advantage that the independence of the observer and its equivalence with Eq.\ \eqref{equation:Lorentziandistanceformula1} will be manifest.

\section{Two-spinor algebra and spacetime}

We give a sketchy account of
a partly original approach to spinors and spacetime geometry, studied by the first author~\cite{canarutto98,canarutto00b,canarutto07},
that has some differences with the well-known
Penrose-Rindler formalism~\cite{penrose84,penrose86}.
We start by noting that a finite-dimensional complex vector space $U$ yields
its complex dual and anti-dual spaces,
$U^\lin$ and $\overline{U}{}^\lin$,
respectively constituted of all linear and anti-linear maps \txtm{U\to\mathbb{C}}.
The space \txtm{\overline{U}\equiv(\overline{U}{}^\lin)^\lin} is then called the \emph{conjugate space} of $U$.
Complex conjugation determines anti-isomorphisms
\txtm{U^\lin\leftrightarrow\overline{U}{}^\lin}
and \txtm{U\leftrightarrow\overline{U}}.
In the standard way of writing coordinate expressions,
conjugate spaces typically yield `dotted indices'.

The space $U\,{\otimes}\,\overline{U}$ is endowed with the anti-linear involution
characterized by the rule
\txtm{u\,{\otimes}\,\bar v\mapsto v\,{\otimes}\,\bar u}\,;
accordingly we get its splitting
\[ U\,{\otimes}\,\overline{U}=\mathrm{H}(U\,{\otimes}\,\overline{U})\oplus\mathrm{iH}(U\,{\otimes}\,\overline{U}) \]
into the real subspaces (of the same dimensions) of all \emph{Hermitian} and anti-Hermitian tensors.
A Hermitian 2-form on $U$ is an element in
\txtm{\mathrm{H}(\overline{U}{}^\lin\,{\otimes}\,U^\lin)\cong(\mathrm{H}(U\,{\otimes}\,\overline{U}))^*}.

Now we consider the case when \txtm{\dim U=2} and the 1-dimensional space $\wedge^{\!2}U$
(not $U$) is endowed with a positive Hermitian metric.
This yields, up to a phase factor, a unique normalized `complex symplectic' tensor
\txtm{\upepsilon\in\wedge^{\!2}U^\lin}.
Accordingly one gets `index moving' isomorphisms
\txtm{\upepsilon^\flat:U\to U^\lin:u\mapsto u^\flat\equiv\upepsilon(u,\_)}
and \txtm{\upepsilon^\#:U^\lin\to U:\lambda\mapsto\lambda^\#\equiv
\upepsilon^{-1}(\lambda,\_)}
that are unique up to a phase factor.

The object
\txtm{\eurm{g}\equiv\upepsilon\,{\otimes}\,\bar\upepsilon\in\wedge^{\!2}U^\lin{\otimes}\wedge^{\!2}\overline{U}{}^\lin}
is then well-defined independently of the choice of a normalized $\upepsilon$\,,
and the rule
\txtm{\eurm{g}(u\,{\otimes}\,\bar v,u'\,{\otimes}\,\bar v')\equiv\upepsilon(u,u')\bar\upepsilon(\bar v,\bar v')}
makes it a bilinear form on $U\,{\otimes}\,\overline{U}$\,.
Its restriction to the 4-dimensional real vector space
\[ H\equiv\mathrm{H}(U\,{\otimes}\,\overline{U}) \]
turns out to have signature \txtm{({+}\,{-}\,{-}\,{-})}\,.
Thus $(H,\eurm{g})$ is a Minkowski space.
Isotropic elements in $H$ are of the form $\pm u\,{\otimes}\,\bar u$ with \txtm{u\in U},
so that there is a natural way of fixing a time orientation in $H$\,.

Now the 4-dimensional complex space
\[ W\equiv U\oplus\overline{U}{}^\lin \]
can be identified with the space of Dirac spinors, that may be represented as \txtm{\psi\equiv(u,\bar\lambda)}\,.
In fact the linear map \txtm{\upgamma:U\,{\otimes}\,\overline{U}\to\mathrm{End}(W)} characterized by
\[ \upgamma[r\,{\otimes}\,\bar s](u,\bar\lambda)\equiv
\sqrt2\,\bigl(\langle\bar\lambda,\bar s\rangle\,r\,,\,\upepsilon(r,u)\,\bar\upepsilon^\flat(\bar s)\bigr) \]
restricts to a Clifford map \txtm{H\to\mathrm{End}(W)}.
Its image generates the \emph{Dirac algebra}, that has the distinguished element $\gD$\,,
corresponding to the volume form $\upeta$ of $H$,
acting as \txtm{\gD(u,\bar\lambda)=\mathrm{i}\,(u,-\bar\lambda)}
and determining the projections \txtm{\frac12\,(1\mp\mathrm{i}\gD)}
onto the \emph{chiral subspaces} \txtm{U\,{\oplus}\,\{0\}} and \txtm{\{0\}\,{\oplus}\,\overline{U}{}^\lin}.

The \emph{Dirac adjunction} is the anti-linear involution
\[ W\to W^\lin\equiv U^\lin\,{\oplus}\,\overline{U}:(u,\bar\lambda)\mapsto(\lambda,\bar u)~,\]
and is associated with the Hermitian product
\[ \eurm{k}:\overline W\times W\to\mathbb{C}:\bigl((\bar u,\lambda),(v,\bar\mu)\bigr)\mapsto
\langle\lambda,v\rangle+\langle\bar\mu,\bar u\rangle \]
that turns out to have signature \txtm{({+}\,{+}\,{-}\,{-})}\,.

It should be noticed that we assume no \emph{positive} Hermitian metric on $U$\,.
Indeed such object is a timelike, future-oriented element \txtm{\eurm{h}\in H^*},
so that its assignment is essentially equivalent to fixing an observer.
It can always be expressed as
\[ \eurm{h}=\lambda\,{\otimes}\,\bar\lambda+\mu\,{\otimes}\,\bar\mu~,\qquad
\lambda,\mu\in U^\lin~.\]

Similarly a timelike, future-oriented vector can be written as
\[ {\scriptstyle Y}=\tfrac1{\surd2}\,\bigl(
u\,{\otimes}\,\bar u+\lambda^\#\,{\otimes}\,\bar\lambda^\#\bigr)~,\qquad
u\in U\,,~\lambda\in U^\lin~.\]
This particular combination fulfills
\txtm{\eurm{g}({\scriptstyle Y},{\scriptstyle Y})=|\langle\lambda,u\rangle|^2}\,,
and we get
\[ \upgamma[{\scriptstyle Y}](u,\bar\lambda)=
\Bigl(\langle\bar\lambda,\bar u\rangle\,u~,
~\langle\lambda,u\rangle\,\bar\lambda\Bigr)~.\]
Thus the  Dirac spinor \txtm{\psi\equiv(u,\bar\lambda)} fulfills
\txtm{\upgamma[{\scriptstyle Y}]\psi=\pm m\,\psi}
if and only if \txtm{\langle\lambda,u\rangle=\pm m}\,.

The standard matricial formalism of Dirac algebra can be recovered by choosing
a basis \txtm{\bigl(\mathsf{z}_1\,,\mathsf{z}_2\bigr)} of $U$ such that
\txtm{\mathsf{z}_1\wedge\mathsf{z}_2} is normalized.
This yields the $\eurm{g}$-orthonormal basis
\[ \bigl(\uptau_\lambda\bigr)\equiv
\bigl(\tfrac1{\surd2}\upsigma_\lambda^{\,{\scriptscriptstyle AA\idot}}\,
\mathsf{z}_{\scriptscriptstyle A}\,{\otimes}\,
\bar{\mathsf{z}}_{\scriptscriptstyle A\idot}\bigr)\subset H \]
(where $\bigl(\upsigma_\lambda\bigr)$ is the $\lambda$-th Pauli matrix),
the \emph{Weyl} and \emph{Dirac bases} (upper indices label elements of dual bases)
\begin{align*}
&\bigl(\upzeta_\alpha\bigr)\equiv
\bigl(\mathsf{z}_1\,,\mathsf{z}_2\,,
-\bar{\mathsf{z}}^1\,,-\bar{\mathsf{z}}^2\bigr)~,\qquad\alpha=1,2,3,4~,
\\[6pt]
&\bigl(\upzeta'_\alpha\bigr)\equiv
\bigl(\tfrac1{\surd2}(\upzeta_1-\upzeta_3)\,,\,
\tfrac1{\surd2}(\upzeta_2-\upzeta_4)\,,\,
\tfrac1{\surd2}(\upzeta_2-\upzeta_4)\,,\,
\tfrac1{\surd2}(\upzeta_2+\upzeta_4)\,\bigr)
\end{align*}

The usual Weyl and Dirac representations of the Dirac algebra are then recovered as
the matrices of the endomorphisms
\txtm{\upgamma_\lambda\equiv\upgamma[\uptau_\lambda]}\,.

\begin{center}{*}~{*}~{*}\end{center}

Let now \txtm{U\to M} be a 2-spinor bundle over the 4-dimensional basis manifold $M$\,;
then, according to the above constructions,
we get bundles \txtm{H\to M} and \txtm{W\to M}
with induced fiber algebraic structures.
A \emph{tetrad-affine approach} to spacetime geometry can be formulated by
describing the gravitational field as a couple $(\upomega,\uptheta)$\,,
where $\upomega$ is a linear connection of \txtm{U\to M} (\emph{spinor connection})
and \txtm{\uptheta:\mathrm{T}M\to H} is a \emph{soldering form} (or \emph{tetrad}),
i.e.\ a linear invertible fibered morphism; this yields a Lorentz metric
\txtm{\uptheta^*\eurm{g}} on $M$
(still denoted as $\eurm{g}$ if no confusion arises).
The condition \txtm{\nabla\uptheta=0} then determines a metric (possibly non-symmetric)
connection of \txtm{\mathrm{T}M\to M}\,,
and one eventually gets all the geometric structures needed for the theory
of Einstein-Cartan-Maxwell-Dirac fields~\cite{canarutto98}.

\section{Two-spinors and Lorentzian distance}

Let us come to the spectral triple reformulation of the Lorentzian distance formula.
The spectral triple's elements in this case are \cite{franco13}
\begin{enumerate}
\item[$\bullet$]
The algebra \txtm{\mathcal{A}\equiv\mathrm{C}^1(M,\mathbb{R})} with pointwise multiplication.
\item[$\bullet$]
The Hilbert space \txtm{\mathcal{H}\equiv\mathrm{L}^2(M,W)}
of square integrable sections of the spinor bundle \txtm{W\to M}\,,
associated with the choice of a positive Hermitian metric $\eurm{h}$ on $W$.
\item[$\bullet$]
The Dirac operator
\txtm{\nasl\equiv-\mathrm{i}\,\uptheta^a_\lambda\,\upgamma^\lambda\nabla\!_a\equiv-\mathrm{i}\,\upgamma^a\nabla\!_a}
associated with the spin connection.
\end{enumerate}
Moreover one uses
\begin{enumerate}
\item[$\bullet$]
The endomorphism $\mathrm{i}\upgamma^0$,  referred in~\cite{franco13} as the {\em fundamental symmetry}.
\item[$\bullet$]
The chirality operator \txtm{\chi\equiv\pm\mathrm{i}\gD=\pm\mathrm{i}\upgamma_0\upgamma_1\upgamma_2\upgamma_3}\,.
\end{enumerate}

Taking different notations and conventions into account|in particular opposite metric signatures|we can rewrite
the Lorentzian distance in the spectral triple formulation as presented by  Franco and Eckstein as
\begin{equation}\tag{L2}\label{equation:Lorentziandistanceformula2}
\dLor(x,y)=\inf_{f\in\mathcal{A}}\Bigl\{
[f(x)-f(y)]^+:\eurm{h}\bigl(\bar\psi\,,\,\upgamma^0([\nasl,f]\pm\gD)\psi\bigr)\geq0, \ \forall\psi\in\mathcal{H}\Bigr\}~.
\end{equation}
We wish to show, through the elaboration of the right-hand side, that this equation is equivalent to the Lorentzian distance formula. Since the latter has been proved, this equation will also be proved.

We observe that the above use of $\upgamma^0$ implies the choice of an observer
(identified with the element $\uptau_0$ of the used Pauli frame),
and that the positive Hermitian metric $\eurm{h}$
depends on the choice of an observer, too.
It is then natural to assume that these are the same observer;
in that case we have \txtm{\eurm{h}(\bar\psi,\upgamma^0\phi)=\eurm{k}(\bar\psi,\phi)},
where $\eurm{k}$ is the (observer-independent) Dirac adjuction metric.
Thus we obtain the more satisfactory version
\begin{equation} \tag{L2'}\label{equation:Lorentziandistanceformula3}
\dLor(x,y)=\inf_{f\in\mathcal{A}}\Bigl\{
[f(x)-f(y)]^+:\eurm{k}\bigl(\bar\psi\,,\,([\nasl,f]\pm\gD)\psi\bigr)\geq0, \ \forall\psi\in\mathcal{H}\Bigr\}~.
\end{equation}
In this expression the observer is still present though, since it is needed in the
choice of $\eurm{h}$\,, which is used in the definition of $\mathcal{H}$\,.
By the way, we note that the Hilbert spaces associated with different observers
may not coincide, since a given section can be square-integrable
with respect to an observer and not with respect to the other
(as it can be easily shown by a suitable counterexample).

We have
\begin{align*}
[\nasl,f]\psi&=\nasl(f\psi)-f\nasl\psi=\upgamma^a\nabla\!_a(f\psi)-f\nasl\psi
=(\upgamma^a\partial_af)\,\psi=\upgamma^\#[\mathrm{d} f]\psi\equiv\dsl\!f\psi~,
\end{align*}
whence for \txtm{\psi\equiv(u,\bar\lambda):M\to U\oplus\overline{U}^\lin\equiv W}
we get
\begin{align*}
&\eurm{k}\bigl(\bar\psi\,,\,([\nasl,f]\pm\gD)\psi\bigr)=
\eurm{k}\bigl(\bar\psi\,,\,(\dsl\!f\pm\gD)\psi\bigr)
=\mp2\,|\langle\lambda,u\rangle|\,\sin\alpha
+\sqrt2\,\bigl\langle\cev\uptheta{}^*(\mathrm{d}\!f)\,,\,u\,{\otimes}\,\bar u+(\lambda\,{\otimes}\,\bar\lambda)^\#\bigr\rangle\bigr)~,
\end{align*}
where \txtm{\alpha\equiv\arg\langle\lambda,u\rangle} that is
\[ \sin\alpha=\frac{\langle\lambda,u\rangle-\langle\bar\lambda,\bar u\rangle}{2\,\mathrm{i}\,|\langle\lambda,u\rangle|}~.\]
Now setting
\[{\scriptstyle Y}\equiv\tfrac{1}{\surd2}\,(u\,{\otimes}\,\bar u+\lambda^\flat\,{\otimes}\,\bar\lambda^\flat)~,\qquad
|{\scriptstyle Y}|^2\equiv\eurm{g}({\scriptstyle Y},{\scriptstyle Y})=|\langle\lambda,u\rangle|^2~,\]
we get
\[\eurm{k}\bigl(\bar\psi\,,\,(\dsl\!f\pm\gD)\psi\bigr)=
2\,\langle\mathrm{d}\!f,{\scriptstyle Y}\rangle\mp2\,|{\scriptstyle Y}|\,\sin\alpha~.\]

At each spacetime point we obtain an arbitrary timelike future-pointing vector
\txtm{{\scriptstyle Y}\in H} by the above expression
(this is essentially the contravariant 4-momentum
associated with an appropriate Dirac spinor
\txtm{\psi\equiv(u,\bar\lambda)}).
Thus \eqref{equation:Lorentziandistanceformula2} can be equivalently written as
\[ \dLor(x,y)=\inf_{f\in\mathcal{A}}\Bigl\{
[f(x)-f(y)]^+:\langle\mathrm{d}\!f,{\scriptstyle Y}\rangle\mp2\,|{\scriptstyle Y}|\,\sin\alpha\geq0
~~\forall {\scriptstyle Y}\in I^+,~\alpha\in\mathbb{R}\Bigr\}~,\]
where \txtm{I^+\subset H} denotes the interior of the future cone.
Moreover we observe that, since $\alpha$ is real,
for any \txtm{{\scriptstyle Y}\in I^+} both conditions
\begin{align*}
&\langle\mathrm{d}\!f,{\scriptstyle Y}\rangle\geq|{\scriptstyle Y}|\,\sin\alpha\quad\forall\alpha\in\mathbb{R}~,
\\[6pt]
&\langle\mathrm{d}\!f,{\scriptstyle Y}\rangle\geq-|{\scriptstyle Y}|\,\sin\alpha\quad\forall\alpha\in\mathbb{R}~,
\end{align*}
are equivalent to
\[ \langle\mathrm{d}\!f,{\scriptstyle Y}\rangle\geq|{\scriptstyle Y}|~.\]
Hence eventually we get
\[ \dLor(x,y)=\inf_{f\in\mathcal{A}}\Bigl\{[f(x)-f(y)]^+:\langle\mathrm{d}\!f,{\scriptstyle Y}\rangle \geq |{\scriptstyle Y}|,
~~\forall {\scriptstyle Y}\in I^+\Bigr\}~,\]
which is essentially the expression~\eqref{equation:Lorentziandistanceformula1}
of the Lorentzian distance
proved by the first author.

Finally we remark that, in the above expression, the observer and
the spinor $\psi$ disappeared, as well as any spinor-related objects;
in particular, the condition that $\psi$ belong to a Hilbert space has no role at all.


\def\cprime{$'$}
\providecommand{\newblock}{}

\end{document}